\documentclass{PoS}

\title{The Timing Noise of Magnetars}

\ShortTitle{The Timing Noise of Magnetars}

\author{\speaker{D. \c{C}erri-Serim}\\
        Middle East Technical University\\
        E-mail: \email{danjela@astroa.physics.metu.edu.tr}}
        
\author{M. M. Serim\\
        Middle East Technical University\\
        E-mail: \email{muhammed@astroa.physics.metu.edu.tr}}
\author{D. Y\"{u}calan\\
       Middle East Technical University \\
        E-mail: \email{yucalan.doga@metu.edu.tr}}        
\author{\c{S}. \c{S}ahiner\\
       Middle East Technical University \\
        E-mail: \email{seyda@astroa.physics.metu.edu.tr}}
\author{S. \c{C}. \.{I}nam\\
        Ba\c{s}kent University\\
        E-mail: \email{inam@baskent.edu.tr}}
\author{A. Baykal\\
        Middle East Technical University\\
        E-mail: \email{altan@astroa.physics.metu.edu.tr}}

\abstract{We represent noise strength analysis of Anomalous X-Ray Pulsars (AXPs) 4U 0142+61, 1RXS J170849.9-400910, 1E 1841-045, 1E 2259+586 and Soft Gamma Repeaters (SGRs) SGR J1833-0832, SWIFT J1822.3-1606 and SWIFT J1834.9-0846 together with the X-Ray binaries  GX 1+4 and 4U 1907+09 for comparison with accreting sources. Using our timing solutions, we extracted residuals of pulse arrival times after removal of spin down trends and we calculated assoicated noise strength of each source. Our preliminary results indicate that the noise strength  is scaling up with spin-down rate. This indicates that, increase in spin-down rate leads to more torque noise on the magnetars. In addition, we present our analysis with Bayesian statistics on the previously reported transient QPO feature of 4U 1907+09. }

\FullConference{11th INTEGRAL Conference Gamma-Ray Astrophysics in Multi-Wavelength Perspective,\\
		10-14 October 2016\\
		Amsterdam, The Netherlands}

\begin{document}

\section{Introduction}
Magnetars are isolated neutron stars with extreme magnetic fields (typically higher than $10^{14}G$) (Kouveliotou et al. 1998; Vasisht \& Gotthelf 1997) and their emission is thought to be originating from decay of this field (Duncan \&Thompson 1992;Beloborodov 2009).
They exhibit short X-ray bursts arising from crustal movements and magnetic activity (Thompson et al. 2002).
Magnetars occasionally show glitches in their spin frequency followed by flux enchancements and structural changes in pulse profiles (Dib \& Kaspi 2014; Rea \& Esposito 2011). 
Timing studies on magnetars reveal that they posses high timing noise level compared with standard pulsars (Woods et al. 2002; Gavriil \& Kaspi 2002). 
In this study, we present our timing noise strength measurements on 7 magnetars.

Pulsars in X-ray binary systems are powered by accretion of matter from
their companion stars (Pringle \& Rees 1972). Angular momentum content of
accreted matter is a source of external torque for the pulsar and causes
intrinsic variations in spin period. Investigating fluctuations in torque
provides information about accreting matter and the internal structure of
the pulsar (Lamb et al. 1978).

\section{Analysis and Noise Strengths}
\begin{figure}[b]
\centering
\includegraphics[width=0.75\linewidth]{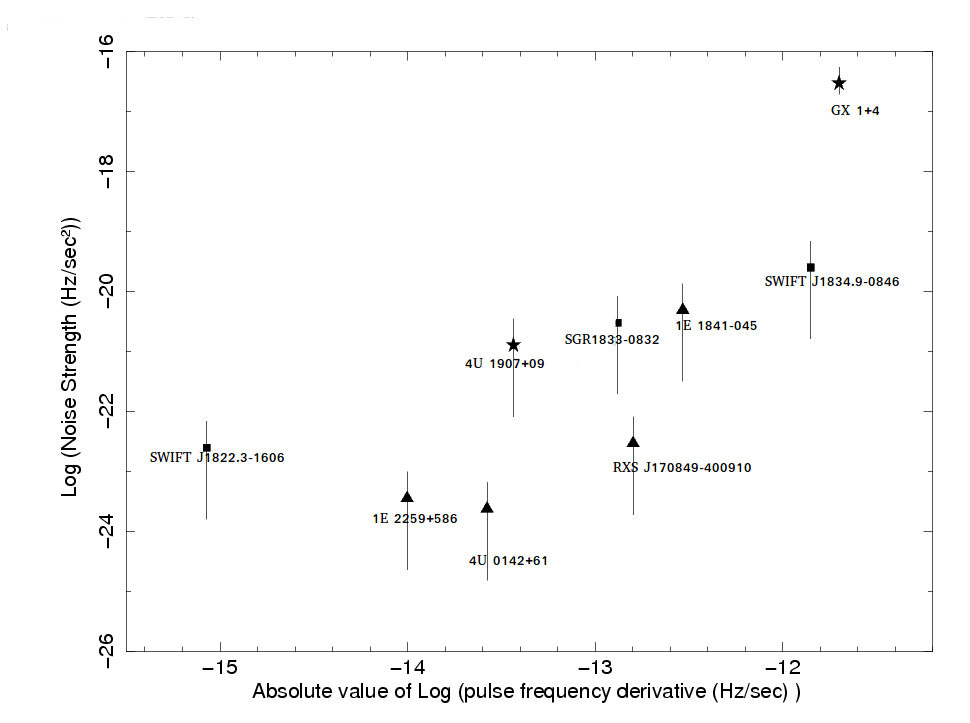} 
\caption{Noise strengths of 7 magnetars and 2 X-ray binaries as a function of their pulse frequency derivatives. Square, triangle and asteriks marks indicate SGRs, AXPs and X-Ray Binaries, respectively.}
\end{figure}
We measured pulse frequencies via phase coherent timing solutions for each source. 
The constructed pulse frequency histories have approximately same time span.  
In order to investigate the torque fluctuations of these sources, we calculated associated noise strengths for the pulse frequency variations (see Cordes 1980;Deeter \& Boynton 1982  for approach and also Baykal 1997 for applications).
The expected mean square residual, after removal of m$^{th}$ degree polynomial over a time span \textit{T}, is given by
\begin{equation}
 \langle \sigma_{R}^{2} (m, T)\rangle = S_{r} T^{2r-1} \langle \sigma_{R}^{2} (m, 1)\rangle_{e}
\end{equation}
where $\langle \sigma_{R}^{2} (m, 1)\rangle_{e}$ indicates the expectation value for unit-strength noise process.
The numerical evaluation of these values are presented in Deeter (1984).
We removed spin down trends from the frequency measurements and calculated corresponding noise strength $S_{r}$ from the mean square residuals. 
In this work, we represent noise strength estimations of four AXPs (1E 2259+586, AXPs 4U 0412+61, 1RXS J170849.9-400910	and 1E 1841-045) and two X-ray binaries (GX 1+4:Serim et al. (2017); Bildsten et al. (1997) and 4U 1907+09: \c{S}ahiner et al. 2012) in addition to three SGRs (SWIFT J1822.3-1606, SGR J1833-0832 and SWIFT J1834-0846; Serim et al. 2012) .
Noise strengths versus absolute value of pulse frequency derivatives are presented in Figure 1.

\section{QPO of 4U 1907+09}
4U 1907+09 is an X-ray pulsar with a spin period of 441.8 s. The secular
spin down trend of the pulsar was interrupted by a torque reversal event in
2004 May (Fritz et al. 2006). After 2007 June, the pulsar returned to spin
down with a rate close to its historical steady spin down rate (\.{I}nam et al.
2009, \c{S}ahiner et al. 2012). 4U 1907+09 is known to show abrupt variations
in X-ray flux such as flares and dips on time scales of minutes to hours
(In't Zand et al. 1997). Quasi-periodic oscillations (QPO) of about 18 s had been
reported during a 1 hr flare on 1996 February 23, which was interpreted as
a transient accretion disc formation (In't Zand et al. 1998).

We re-analysed the RXTE-PCA observation taken on February 23, 1996. We extracted the lightcurve of the observation and searched for QPO with the Bayesian search algorithm written in
Python by Huppenkothen\footnote{The associated Python codes can be found in https://github.com/dhuppenkothen/BayesPSD}.
\begin{figure}
\centering
\includegraphics[width=0.4\linewidth]{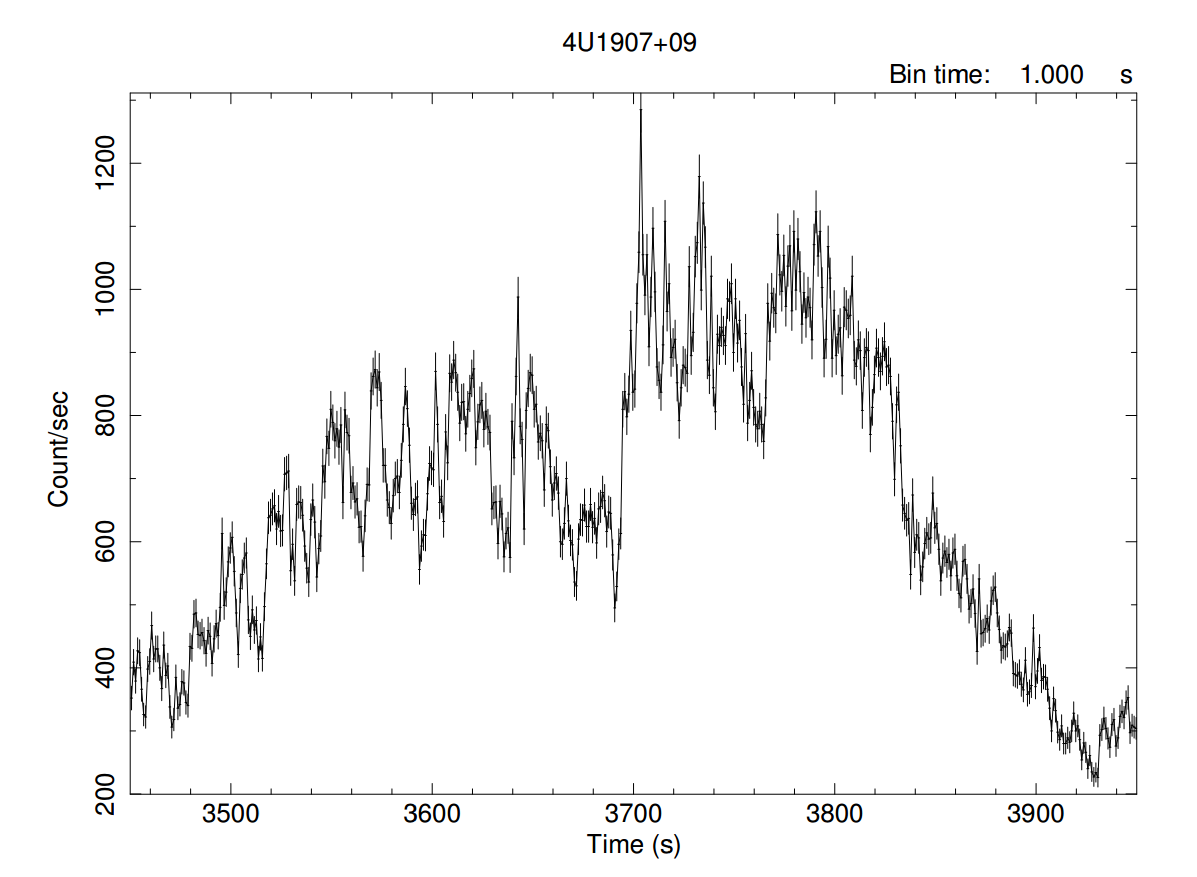}
\includegraphics[width=0.43\linewidth]{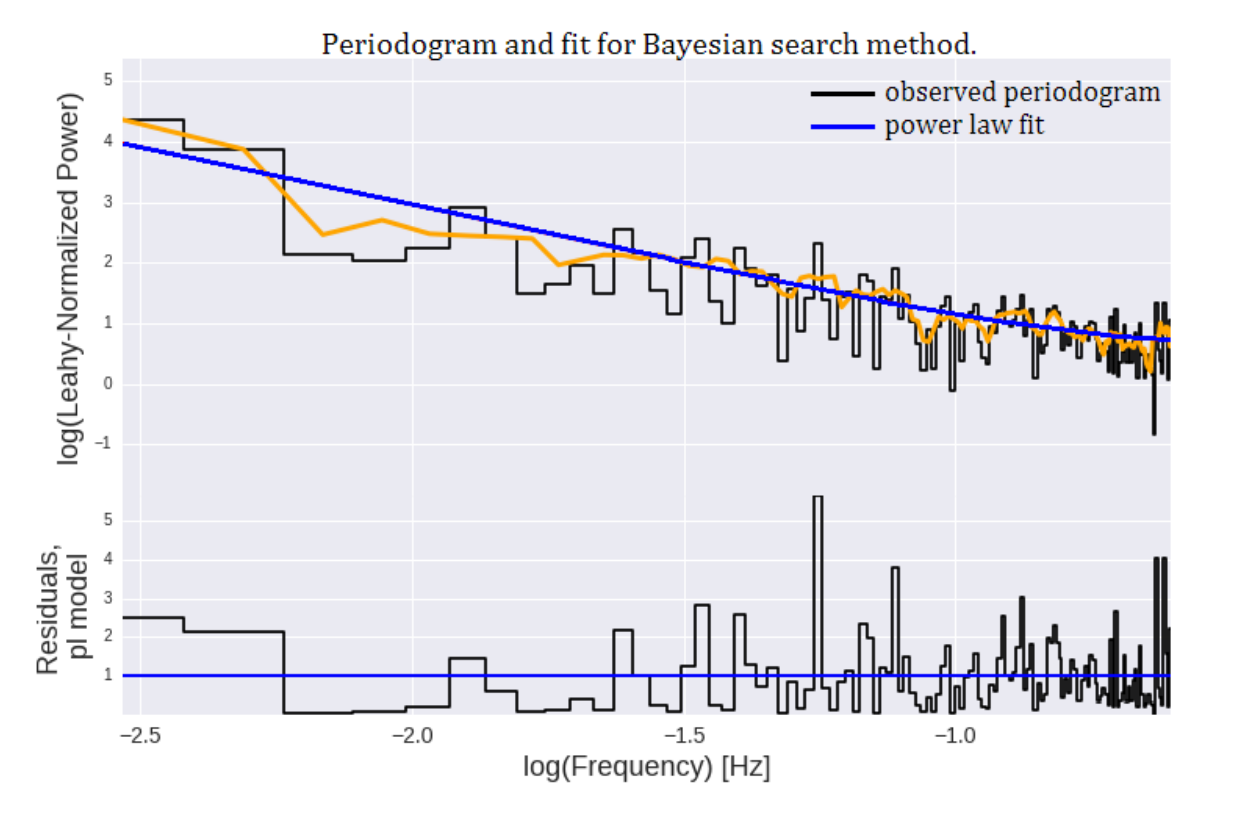}  
\caption{ Lightcurve of 4U 1907+09 February 23, 1996 observation (left) and the periodogram and the power law fit result for the Bayesian
search analysis (right).}
\end{figure}
The Bayesian search method found a QPO candidate
around 17.9 s. The probability of detecting a false signal is
p($\chi^{2}$ > 11.277) = $3.6 \times 10^{-3}$ 
, which corresponds to a 2.91$\sigma$ level of detection. We also constructed power density spectrum (PDS) of the same lightcurve, then fit the continuum with power law model and calculated the significance of qpo signal. In this method, detection level of the signal yielded 3.54$\sigma$.
\section{Conclusion}
In conclusion, we extract the timing residuals of 7 magnetars (3 SGRs and 4 AXPs) and 2 X-ray binaries using RXTE and Swift observations and calculated associated timing noise strengths. 
 
\begin{itemize}
 \item Our pulse timing noise analysis on 7 magnetars indicate that the spin-down rates of magnetars seem to show a correlation with their noise strengths. These noise strengths could be the result of micro-scale cracking of the 
crust due to the magnetic field stress on the crust.  We will extend this work to other magnetars listed in McGill Magnetar Catalog (Olausen \& Kaspi 2014).

\item We re-analysed the transient QPO feature of 4U 1907+09. The Bayesian method does find a candidate for a QPO around $\simeq$18s, however, lowers the level of detection to 2.91$\sigma$.
The level of detection is less than that of the comparison
analysis, which finds a candidate around the same period, with 3.54. 
\end{itemize}
We acknowledge support from T\"{U}B\.{I}TAK, the Scientific and Technological Research Council of Turkey through the research project MFAG 114F345 and Scientific Research Projects Coordination Unit (BAP) with the project BAP-01-05-2016-003.


\begin{thebibliography}{99}

\bibitem{1997A&A...319..515B}{{Baykal}, A.}, {\it The torque and X-ray flux changes of OAO 1657-415}, AAP 318, 515 (1997)

\bibitem{2009ApJ...703.1044B}{{Beloborodov}, A.~M.}, {\it Untwisting Magnetospheres of Neutron Stars}, ApJ 703, 1044 (2009)

\bibitem{1997ApJS..113..367B}{Bildsten}, L. et al., {\it Observations of Accreting Pulsars}, ApJS 113, 367 (1997)

\bibitem{1980ApJ...237..216C}{Cordes}, J.~M., {\it Pulsar timing. II - Analysis of random walk timing noise - Application to the Crab pulsar}, ApJ 237, 216 (1980)

\bibitem{1982ApJ...261..337D}{{Deeter}, J.~E. and {Boynton}, P.~E.}, {\it Techniques for the estimation of red power spectra. I - Context and methodology}, ApJ 261, 337 (1982)

\bibitem{1984ApJ...281..482D}{{Deeter}, J.~E.}, {\it Techniques for the estimation of red power spectra. II Evaluation of alternative methods}, ApJ 281, 482 (1984)

\bibitem{2014ApJ...784...37D}{{Dib}, R. and {Kaspi}, V.~M.}, {\it 16 yr of RXTE Monitoring of Five Anomalous X-Ray Pulsars}, ApJ 784, 37 (2014) 

\bibitem{1992ApJ...392L...9D}{{Duncan}, R.~C. and {Thompson}, C.}, {\it Formation of very strongly magnetized neutron stars - Implications for gamma-ray bursts}, ApJL 392, 9 (1992) 

\bibitem{2006A&A...458..885F}{Fritz}, S. et al., {\it A torque reversal of 4U 1907+09}, AAP 358, 885 (2006)

\bibitem{2002ApJ...567.1067G}{{Gavriil}, F.~P. and {Kaspi}, V.~M.}, {\it Long-Term Rossi X-Ray Timing Explorer Monitoring of Anomalous X-Ray Pulsars}, ApJ 567, 1067 (2002)

\bibitem{2009MNRAS.395.1015I}{{Inam}, S.~{\c C}. and {{\c S}ahiner}, {\c S}. and {Baykal}, A.}, {\it Recent torque reversal of 4U1907+09}, MNRAS 395, 1015 (2009)

\bibitem{1997ApJ...479L..47I}{{in 't Zand}, J.~J.~M. and {Strohmayer}, T.~E. and {Baykal}, A.}, ApJL 479, 47 (1997)

\bibitem{1998ApJ...496..386I}{{in 't Zand}, J.~J.~M. and {Baykal}, A. and {Strohmayer}, T.~E.}, {\it Recent X-Ray Measurements of the Accretion-powered Pulsar 4U 1907+09}, ApJ 496, 386 (1998)

\bibitem{1978ApJ...225..582L}{{Lamb}, F.~K. and {Pines}, D. and {Shaham}, J.}, {\it Period variations in pulsating X-ray sources. II - Torque variations and stellar response}, ApJ 225, 582 (1978)

\bibitem{1998Natur.393..235K}{Kouveliotou}, C. et al., {\it An X-ray pulsar with a superstrong magnetic field in the soft {$\gamma$}-ray repeater SGR1806 - 20}, Nature 393, 237 (1998)

\bibitem{2014ApJS..212....6O}{{Olausen}, S.~A. and {Kaspi}, V.~M.}, {\it The McGill Magnetar Catalog},  ApJS 212, 6 (2014)

\bibitem{1972A&A....21....1P}{{Pringle}, J.~E. and {Rees}, M.~J.}, {\it Accretion Disc Models for Compact X-Ray Sources}, AAP 21, 1 (1972)

\bibitem{2011ApJ...729L..21R}{Rea}, N. et al.,{\it The X-ray Quiescence of Swift J195509.6+261406 (GRB 070610): An Optical Bursting X-ray Binary?}, ApJL 729, 21 (2011)

\bibitem{2012ASPC..466..255S}{{Serim}, M.~M. and {Inam}, S.~{\c C}. and {Baykal}, A.}, {\it Noise Strength Estimates of Three SGRs: Swift J1822.3-1606, SGR J1833-0832 and Swift J1834.9-0846}, ASPC 466, 255 (2012)

\bibitem{asd} Serim et al., {\it Comprehensive Timing and Spectral Analysis of GX 1+4} to be submitted to MNRAS

\bibitem{2012MNRAS.421.2079S}{{{\c S}ahiner}, {\c S}. and {Inam}, S.~{\c C}. and {Baykal}, A.}, {\it A comprehensive study of RXTE and INTEGRAL observations of the X-ray pulsar 4U 1907+09}, MNRAS 421, 2079 (2012)

\bibitem{2002ApJ...574..332T}{{Thompson}, C. and {Lyutikov}, M. and {Kulkarni}, S.~R.}, {\it Electrodynamics of Magnetars: Implications for the Persistent X-Ray Emission and Spin-down of the Soft Gamma Repeaters and Anomalous X-Ray Pulsars}, ApJ 574, 332 (2002)

\bibitem{1997ApJ...486L.129V}{{Vasisht}, G. and {Gotthelf}, E.~V.}, {\it The Discovery of an Anomalous X-Ray Pulsar in the Supernova Remnant Kes 73}, ApJL 486, 129 (1997)

\bibitem{2002ApJ...576..381W}{Woods}, P.~M. et al., {\it Large Torque Variations in Two Soft Gamma Repeaters}, ApJ 576, 381 (2002)

\end{thebibliography}
\end{document}